\documentstyle[epsfig]{elsart}
\begin{document}
\runauthor{Ishida}
\begin{frontmatter}
\title{Covariant Classification of $q\bar q$-Meson Systems
and Existence of New Scalar and Axial-Vector Mesons}
\author[NU]{Shin Ishida}
\address[NU]{ Atomic Energy Research Institute, 
College of Science and Technology,\\
Nihon University, Tokyo 101-0062, Japan}
\author[TIT]{Muneyuki Ishida}
\address[TIT]{Department of Physics, Tokyo Institute of Technology,
Tokyo 152-8551, Japan}
\begin{abstract}
The level classification of quark-antiquark systems is 
generally made by the $LS$-coupling scheme, resorting to the
non-relativistic quark model. However, it has been well known
that the $\pi$-meson with the 
$(L,S)=(0,0)$ shows also the properties as a 
Nambu-Goldstone boson in the case of spontaneous
breaking of a relativistic symmetry, the chiral symmetry.  
In this talk I present a covariant classification scheme for 
describing both the non-relativistic and the relativistic 
$q\bar q$-mesons and point out possibility for existence of 
the new scalar meson nonet (to be assigned to the $\sigma$ nonet)
and also of the new axial-vector meson nonet, ``chiralons,'' 
which should be 
discriminated from the conventional $^3P_0$ and $^3P_1$ states,
respectively.
\end{abstract}
\end{frontmatter}
%
\begin{center}
{\bf [Introduction]}
\end{center}

\vspace*{-0.3cm}

Recently strong evidences for existence of the light $\sigma (600)$
meson have been given by reanalyzing\cite{ref1} the old $\pi\pi$ 
scattering phase shift data. The evidences\cite{ref1} have been also
given in the production processes. Reflecting these situations the
$\sigma$ has been revived in the lists of PDG'96 and '98 after 
missing over two decades. Furthermore, some evidences for existence of
the $I=1/2$ scalar meson $\kappa (900)$ have been also reported\footnote{
We have pointed out a possibility that these resonances together with 
the $a_0(980)$ and the $f_0(980)$ belong to the $\sigma$-nonet in the
$SU(3)$ linear $\sigma$ model. See, M. Y. Ishida in 
this workshop\cite{ref2}. 
}
by reanalyzing\cite{ref1} the $K\pi$ scattering phase shifts.
These $\sigma (600)$ and $\kappa (900)$ mesons seem to be out of the
conventional PDG-classification scheme based on the non-relativistic
quark model(NRQM) 
and considered to be discriminated from the $q\bar q$-$^3P_0$
states with the higher masses.
As is wellknown, there are the two contrasting viewpoints of 
$q\bar q$-mesons(see Table I): the one is non-relativistic, based on the
approximate symmetry of the $LS$-coupling in NRQM; 
while the other one is relativistic, based on the 
dynamically broken chiral symmetry in the NJL model.
\begin{table}
\begin{center}
\caption{\hspace*{3cm}Two contrasting viewpoints of $q\bar q$ mesons}
\begin{tabular}{lcc}
\hline
\hline
      & Non-Relativistic & Relativistic \\
\hline
Model & Non R. quark Model & NJL Model  \\
\hline
Approx. Symm. & $LS$ coupling & Chiral symm. \\
\hline
quark mass & $m_q^{\rm const}$ large &  $m_q^{\rm curr}$ small\\
\hline
\end{tabular}
\label{tab:two}
\end{center}
\end{table}
Now it is widely believed that the $\pi$ meson (or $\pi$ nonet)
has a dual nature of a ``non-relativistic'' particle with the 
$(L,S)=(0,0)$ and also of a relativistic particle as a 
Nambu-Goldstone boson with
$J^P=0^-$ in the case of spontaneous chiral symmetry-breaking.
The purpose of this talk is to present a unification
scheme of the above two viewpoints, giving covariant quark spin wave
functions of the $q\bar q$-mesons. 
%
\begin{center}
{\bf [Covariant extension of $LS$ coupling scheme]}
\end{center}

\vspace*{-0.3cm}

({\it Boosted LS coupling scheme})\ \ \ \ For many years we have 
developed the boosted $LS$ coupling(bLS) scheme in the covariant 
oscillator quark model(COQM)\cite{ref3} 
as a covariant extension of the $LS$
coupling scheme in NRQM. The meson wave functions(WF) in COQM
are tensors in $\tilde U(4)\times O(3,1)$ space and reduce 
at the rest frame  to those in the 
$SU(2)_{\rm spin}\times O(3)_{\rm orbit}$-space in NRQM.\\
All $q\overline{q}$-mesons are 
described unifiedly by the bilocal field
${\Phi_A}^B(x_1,x_2),$
where $x_1(x_2)$ is a space-time 
coordinate of a constituent quark (anti-quark), $A=(a, \alpha)$ $(B=(b,\beta))$ 
the flavor and Dirac spinor indices. 
The $\Phi$ fields are supposed to satisfy the bilocal 
Klein-Gordon equation of 
Yukawa with the squared-mass operator, and they are 
decomposed into the Fierz-components(, representing 
respective mesons,), that is, 
eigen states of ${\cal M}^2$  
as 
\begin{eqnarray}
& & \Phi(X, x)=\sum_{P,n}(e^{iP_n\cdot X}\Psi_n(x;P_n)
+e^{-iP_n\cdot X}\bar\Psi_n(x;P_n)) \\
& & {\cal M}^2(x_\mu,\frac{\partial}{\partial x_\mu})\Psi_n(x;P_n)=
M^2_n\Psi_n(x;P_n)
\end{eqnarray}
The internal wave 
functions $\Psi (x;P_n)$ are factorized into the space-time
portion $f_n$ and the spinor one 
$U_n\ (\bar U_n\equiv -\gamma_4U^\dagger\gamma_4)$ as 
$\Psi_n =f_n(x,P_n)U(P)$,
reflecting the general framework of bLS scheme.
In COQM the $f_n$ is expanded in terms of the covariant oscillator
functions and the $U$ is taken as a Bargmann-Wigner spinor.\\
The physical reason for validity of bLS scheme(in treating directly
mass spectra instead of mass itself) comes from a phenomenological 
fact that the quark binding potential is dominantly central: 
It is supported from the following successful results of applications
of COQM: The prediction of the same form factor relations of 
weak semi-leptonic decays in the heavy-light quark system as obtained 
in the heavy-quark effective theory. The derivation of decay spectra of
$B\rightarrow D/D^*l\nu$ in conformity with experiments.
The well description of qualitative features of the radiative 
transitions of heavy quarkonium systems.\\
({\it Covariant spin WF of non-relativistic particle-BW spinor})\ \ \ \ 
The spin WF, $U$, is assumed
  to 
satisfy the Bargmann-Wigner (BW) equation as 
\begin{eqnarray}
(iP\cdot{\gamma}^{(1)}+M{)_\alpha}^{\alpha^\prime}
{U(P)_{\alpha^\prime}}^\beta &=& 0, \ \ \ \ \ 
{U(P)_\alpha}^{\beta^\prime}(-iP\cdot{\gamma}^{(2)}+M
{)_{\beta^\prime}}^\beta = 0,  
\label{eq7}
\end{eqnarray}
where $P_\mu(M)$ is the four-momentum (mass) of the meson. The 
BW spinor is a covariant 
generalization of the Pauli spinor, and 
decomposed into the irreducible components as 
\begin{equation}
{U(P)_A}^B=\frac{1}{2\sqrt{2}}[(-\gamma_5P_s{(P)_a}^b+i\gamma_\mu
V_\mu{(P)_a}^b)(1+\frac{i\gamma_\mu P_\mu}{M}){]_\alpha}^\beta,
\end{equation}
where $P_s(P)$ $(V_\mu(P))$ represents the pseudoscalar (vector) meson local 
field and $V_\mu(P)$ satisfies the Lorentz condition $P_\mu V_\mu=0$. 
The mass of these (ground state) mesons is degenerate and taken 
as a simple sum of quark 
masses: 
$M\equiv m_1+m_2.$
Then Eq.(\ref{eq7}) is easily seen 
to be equivalent to the following ``free" constituent Dirac 
equations\cite{ref3} 
(with a constraint on the momenta $p_i$) 
\begin{eqnarray}
& & (ip_1\cdot{\gamma}^{(1)}+m_1)U(p_1,p_2)=0 \ \ , \ \ 
U(p_1,p_2)(-ip_2\cdot{\gamma}^{(2)}+m_2)=0,\label{eq8} \\
& & p_{1,2\mu}=\kappa_{1,2}P_\mu ,\ \ \kappa_1+\kappa_2=1\ \ 
(\kappa_{1,2}=m_{1,2}/(m_1+m_2)).
\label{eq218}
\end{eqnarray}
The constraint Eq.(\ref{eq218})  
implies that each constituent is in ``parton-like motion", and 
moving with the equal 3-dimensional velocity to that of total meson:
That is,
${\mbox{\boldmath  $v$}}_{1,2} = 
\frac{{\mbox{\boldmath  $p$}}^{(1,2)}}{p_0^{(1,2)}}
=\frac{\kappa_{1,2}{\mbox{\boldmath $P$}}_M}{\kappa_{1,2}P_{M,0}}
=\frac{{\mbox{\boldmath $P$}}_M}{P_{M,0}}
={\mbox{\boldmath $v$}}_M  .$ 
%
\begin{center}
{\bf [Covariant spin WF of relativistic particle]}
\end{center}

\vspace*{-0.3cm}

({\it Covariant framework for description of $q\bar q$ meson system})
\ \ \ 
Our physical picture of the Yukawa
bilocal function is given by the (B.S.) amplitude
\begin{eqnarray}
{\Phi_A}^B(x_1,x_2) & \approx & 
\langle 0|\psi_A(x_1)\bar\psi^B(x_2)|M\rangle .
\end{eqnarray}
Concerning space-time dependence the amplitude is Fourier-expanded as
 \begin{eqnarray}
{\Phi_A}^B(x_1,x_2) & = & 
\sum_{p_{1\mu},p_{2\mu}} e^{ip_1x_1}e^{ip_2x_2} {\Phi_A}^B(p_1,p_2) 
\nonumber\\
 &=&
\sum_{P_\mu =p_{1\mu}+p_{2\mu}} ( e^{iPX}{\Psi_A}^B(x;P)+
  e^{-iPX}{\bar\Psi_A}^B(x;P)),
\end{eqnarray}
where in the last side is assumed that $p_{i,0}>0$ and 
$P_0=p_{1,0}+p_{2,0}>0.$\\
Concerning spin dependence the amplitude is\footnote{
From here we shall consider only the ground states, or the
``relativistic $S$-wave'' states, neglecting the relative coordinates
$x_\mu$.
} expanded by 
free bi-Dirac spinors $\stackrel{(i)}{W}(P)$ of constituent 
quarks and antiquarks as ($\langle A\rangle$ denoting trace of $A$)
\begin{eqnarray}
{\Psi_\alpha}^\beta (x,P) &=& \sum_i {W_\alpha^{(i)\beta}}(P)
{M}^{(i)}(x,P),\ \ {M}^{(i)}(x,P)\equiv
\langle {W}^{(i)}(P)\Psi (x,P) \rangle ,
\label{eq13}
\end{eqnarray}
and $\bar\Psi_\alpha^\beta$ is similarly represented by 
${\overline{W}_\alpha^{(i)\beta}} $ and 
${M^{(i)\dagger} } =\langle{\overline{W}^{(i)}}
\bar\Psi\rangle .$ \\
({\it Covariant spin WF of relativistic particle})\ \ \ \ 
First let's define the 
conventional positive and negative energy Dirac spinor by
\begin{eqnarray}
\psi_\alpha (x) &=& u_\alpha (p)e^{ipx},\ \  v_\alpha (p)e^{-ipx};\ \ 
p_\mu =({\bf p},iE),\ \ E\equiv\sqrt{{\bf p}^2+m^2}\nonumber\\
(ip\gamma &+& m)u(p)=0,\ \ (ip\gamma -m)v(p)=0. 
\end{eqnarray}
As is easily seen from Eqs. (\ref{eq8}) and (\ref{eq218}), 
the bi-Dirac spinor for the non-relativistic 
particle is given by
\begin{eqnarray}
{W_\alpha^{(NR)\beta} }(P) &\equiv & U_\alpha^\beta (P)
=u_\alpha (p_1)\bar v^\beta (p_2)|_{p_i=\kappa_iP}.
\end{eqnarray}
On the othe hand, we choose as the bi-Dirac spinor 
for the relativistic particle
\begin{eqnarray}
{W_\alpha^{(R)\beta} }(P) &\equiv & C_\alpha^\beta (P)
=u_\alpha (p_1)\bar u^\beta (p_2)|_{p_i=\kappa_iP},
\end{eqnarray}
which satisfies the ``free'' constituent Dirac equations
\begin{eqnarray}
& & (ip_1\cdot{\gamma}^{(1)}+m_1)C(p_1,p_2)=0 \ \ , \ \ 
C(p_1,p_2)(ip_2\cdot{\gamma}^{(2)}+m_2)=0,\label{eq17} \\
& & p_{1,2\mu}=\kappa_{1,2}P_\mu ,\ \ \kappa_1+\kappa_2=1\ \ .
\ \ \ \ \ \ \ \ \ 
\label{eq18}
\end{eqnarray}
These equations (\ref{eq17}) and (\ref{eq18}) are equivalent to the 
new type of BW equation
\begin{eqnarray}
& & (iP\cdot{\gamma}^{(1)}+M)C(P)=0 \ \ , \ \ 
C(P)(iP\cdot{\gamma}^{(2)}+M)=0.
\label{eq19}
\end{eqnarray}
The new BW spinor is decomposed into the irreducible 
components, the scalar $S$ and axial-vector $A_\mu$, as 
\begin{eqnarray}
C(P)_A^B & = & \frac{1}{2\sqrt{2}}\left[\left( 
1-\frac{i\gamma_\mu P_\mu}{M}\right)S(P)_a^b
+\gamma_5\gamma_\mu A_\mu (P)_a^b\left( 
1-\frac{i\gamma_\mu P_\mu}{M}\right)\right]_\alpha^\beta ,
\label{eq20}
\end{eqnarray}
where $A_\mu$ satisfies $P_\mu A_\mu (P)=0 $.\\
Here it is worthwhile to note that both the constituent quark and 
anti-quark, in this case also, are in ``parton-like motion'' and
moving with the equal 3-dimensional velocity to that of total meson.
Especially it is interesting that this situation on the 
anti-quark comes from  
(, noting $\bar u(p_2)\approx\bar v(-p_2),$)
its having anti-parallel motion to the meson and the negative energy,
as\\

\vspace*{-0.5cm}

$
{\mbox{\boldmath $v$}}_2 = 
\frac{{\mbox{\boldmath $p$}}^{(2)}}{\textstyle p_0^{(2)}}
=\frac{\textstyle -\kappa_2{\mbox{\boldmath $P$}}_M}{\textstyle -\kappa_2 P_{0,M}}
=\frac{{\mbox{\boldmath $P$}}_M}{\textstyle P_{0,M}}
={\mbox{\boldmath $v$}}_M.
$
%
\begin{center}
{\bf [Chiral transformation of spin WF]}
\end{center}

\vspace*{-0.3cm}

Since we know the chiral transformation property of the  
bi-Dirac spinors, $W_\alpha^\beta =U_\alpha^\beta$ or $C_\alpha^\beta$,
as
$\ \ 
W  \rightarrow  W'={\rm exp}\ (i\alpha\gamma_5)W
{\rm exp}\ (i\alpha\gamma_5),
\ \ $
we can directly deduce the transformation law 
\begin{eqnarray}
P'_s &=& \ \ {\rm cos}2\alpha P_s+{\rm sin}2\alpha S;\ \ 
\ \ \ \ \ \ V'_\mu =V_\mu\nonumber\\
S' &=& -{\rm sin}2\alpha P_s+{\rm cos}2\alpha S;\ \ 
\ \ \ \ \ \ A'_\mu =A_\mu ,
\end{eqnarray}
for the composite mesons described by
$\ \ 
 P_s  \propto  \langle Ui\gamma_5\rangle ,\ \  
 V_\mu \propto \langle U\gamma_\mu\rangle ,\ \ 
 S  \propto  \langle C\ 1\rangle ,\ \  
 A_\mu \propto \langle Ci\gamma_5\gamma_\mu\rangle .\ \ 
$
Similarly we can derive the conventional transformation laws for mesons,
appearing in any type of linear representations, for example, the 
$SU(2)(SU(3))$ linear $\sigma$ model
\begin{eqnarray}
 & & \stackrel{(i,r)}{\pi}\propto\langle\frac{\tau^{(i)}}{2}
\left(\frac{\lambda^{(r)}}{2}\right) i\gamma_5U\rangle ,\ \ 
\stackrel{(0,r)}{\sigma}\propto\langle \frac{\tau^{(0)}}{2}
\left(\frac{\lambda^{(r)}}{2}\right) C\rangle . 
\end{eqnarray}
The mutual relations of these composite ground state 
mesons are schematically shown in Fig. 1.
\begin{figure}[t]
 \epsfysize=3.6 cm
 \centerline{\epsffile{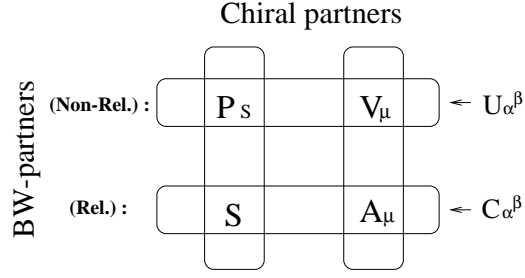}}
 \caption{Mutual relations among the composite ground state mesons 
}
\label{fig:partner}
\end{figure}
\begin{center}
{\bf [Experimental search for chiralons]}
\end{center}

\vspace*{-0.3cm}

In the preceding sections we have shown the possibility of existence
of a new scalar (possibly nonet) and a new axial-vector (possibly nonet),
to be called ``chiralons,'' as the ``relativistic $S$-wave'' states
of composite $q\bar q$ meson systems. Since the former is considered
already to be realized in nature as the $\sigma$-nonet\cite{ref2},
is naturally expected the existence of new axial-vector meson nonet,
$a_{1c}$-nonet, which is to be the BW-partner of $\sigma$-nonet and also
to be the chiral partner of $\rho$-nonet( see Fig. 1).\\
In the symmetric limit the mass of $I=1$ member $a_{1c}$ of the 
$a_{1c}$-nonet is equal to that of $\rho$. In the case of spontaneous
breaking of the symmetry the famous relation had been 
predicted\cite{ref4}.
\begin{eqnarray}
m_{a_{1c}} &=& \sqrt{2}m_\rho \approx 1.1 \ {\rm GeV}.
\end{eqnarray}
The experimental situation in search for the $a_1$ meson seems quite
in confusion such that its mass and width look like to be variant,
depending upon its production and decaying channels.
We infer that this comes from the fact that there exists two
axial-vector mesons with the same quantum numbers, 
$a_1$ and $a_{1c}$.\cite{ref5}\\
It may be important to study experimentally on the existence of 
$a_{1c}$-nonet as well as of $\sigma$-nonet. We consider the 
experimental channels,
\begin{eqnarray}
e^++e^- &\rightarrow & \ \ \phi\ {\rm or}\ J/\psi\ \ \rightarrow\ \ \sigma\ 
{\rm or}\ a_{1c}+\cdots ,\nonumber
\end{eqnarray}

\vspace*{-1cm}

\begin{eqnarray}
\gamma +\gamma\ \  &\rightarrow & \ \ \sigma\ {\rm or}\ a_{1c}+\cdots ,
\end{eqnarray}
in VEPP are promising for this search.

\end{document}